\documentclass{aa}
\usepackage{graphicx}
\usepackage{txfonts}

\begin{document}

\title{Heat transfer in sunspot penumbrae}
\subtitle{Origin of dark-cored penumbral filaments}

\author{B.\ Ruiz Cobo\inst{1} \and L.R.\ Bellot Rubio\inst{2}}
\institute{Instituto de Astrof\'{\i}sica de Canarias, 38200 La Laguna,
Tenerife, Spain  
\and 
Instituto de Astrof\'{\i}sica de Andaluc\'{\i}a, CSIC, Apdo.\ 3004, 
18080 Granada, Spain \\ \email{lbellot@iaa.es}}

\date{Received 13 March 2008 / Accepted 31 May 2008}

\abstract
{Observations at 0\farcs1 have revealed the existence of
dark cores in the bright filaments of sunspot penumbrae. Expectations
are high that such dark-cored filaments are the basic building blocks 
of the penumbra, but their nature remains unknown.}
{We investigate the origin of dark cores in penumbral filaments and 
the surplus brightness of the penumbra. To that end we use an uncombed penumbral model.} 
{The 2D stationary heat transfer equation is solved in a stratified
atmosphere consisting of nearly horizontal magnetic flux tubes embedded
in a stronger and more vertical field. The tubes carry an Evershed
flow of hot plasma.} 
{This model produces bright filaments with dark
cores as a consequence of the higher density of the plasma inside the
tubes, which shifts the surface of optical depth unity toward higher
(cooler) layers. Our calculations suggest that the surplus brightness
of the penumbra is a natural consequence of the Evershed flow, and
that magnetic flux tubes about 250 km in diameter can explain the
morphology of sunspot penumbrae.}
{}

\keywords{sunspots -- Sun: magnetic fields -- Sun:
photosphere -- magnetohydrodynamics (MHD) -- plasmas
\vspace*{1em}  }

\titlerunning{Origin of dark-cored penumbral filaments}
\authorrunning{B.\ Ruiz Cobo and L.R.\ Bellot Rubio}
\maketitle

\section{Introduction}
At high angular resolution, penumbral filaments are
observed to consist of a central dark lane and two lateral
brightenings (Scharmer et al.\ 2002; S\"utterlin et al.\ 2004; Rouppe
van der Voort et al.\ 2004; Bellot Rubio et al.\ 2005; Langhans et
al.\ 2007). The common occurrence of dark-cored filaments and the fact
that their various parts show a coherent behavior have raised
expectations that they could be the fundamental constituents of the
penumbra. Their nature, however, remains enigmatic.

One possibility is that dark-cored filaments represent magnetic flux
tubes carrying a hot flow. This would support the uncombed model
proposed by Solanki \& Montavon (1993), which describes the penumbra
as a collection of nearly horizontal flux tubes embedded in a more
vertical background field. The uncombed model is, by far, the most
successful representation of the fine structure of the penumbra
currently available. It explains the polarization profiles of visible
and near-infrared lines observed in sunspots (e.g., Beck 2008),
including their net circular polarization (NCP). This success is not
trivial, since the behavior of the NCP depends on the details of the
magnetic and velocity fields in a very subtle way (see M\"uller et
al.\ 2002; Borrero et al.\ 2007; Tritschler et al.\ 2007, and
references therein). The uncombed model is supported not only by
observations, but also by theoretical work. Schlichenmaeier et al.\
(1998) and Schlichenmaier (2002) performed numerical simulations of
penumbral flux tubes in the thin tube approximation. These
calculations show filaments whose morphology and dynamics are very
similar to those actually observed in the penumbra (Schlichenmaier
2003). Moreover, the simulations offer a natural explanation for the
Evershed flow, the most conspicuous dynamical phenomenon of
sunspots. In spite of these achievements, it is still not known
whether magnetic flux tubes can also account for the existence of dark
cores in penumbral filaments and, more importantly, for the surplus
brightness of the penumbra. Schlichenmaier \& Solanki (2003) suggested
that hot upflows along magnetic flux tubes would indeed be able to
heat the penumbra to the required degree if the tubes return to the
solar interior after they have released their energy in the
photosphere. At that time the existence of opposite-polarity field
lines in the penumbra was unclear, but now it is a well-established
observational fact: submerging flux tubes have been detected from
Stokes inversions and even imaged directly by Hinode (Sainz Dalda \&
Bellot Rubio 2008). Thus, the Evershed flow remains the best candidate
to explain the brightness of the penumbra.

Another possibility is that the dark-cored filaments are the
manifestation of field-free gaps that pierce the sunspot magnetic
field from below. The concept of a gappy penumbra was proposed by
Spruit \& Scharmer (2006) and Scharmer \& Spruit (2006) as an
alternative way to explain the surplus brightness of the penumbra, on
the assumption that the Evershed flow is not sufficient. The gaps
would sustain normal convection, carrying heat to the solar
surface. Radiative transfer calculations need to be performed to
show that a penumbra consisting of field-free gaps is able to explain
the corpus of spectropolarimetric observations accumulated over the
years. In its present form, however, the gappy model is bound to
experience substantial difficulties when confronted with the
observations (Bellot Rubio 2007).

\begin{figure*}[t]
\begin{center}
\resizebox{.39\hsize}{!}{\includegraphics[bb=-8 8 275 191,clip]{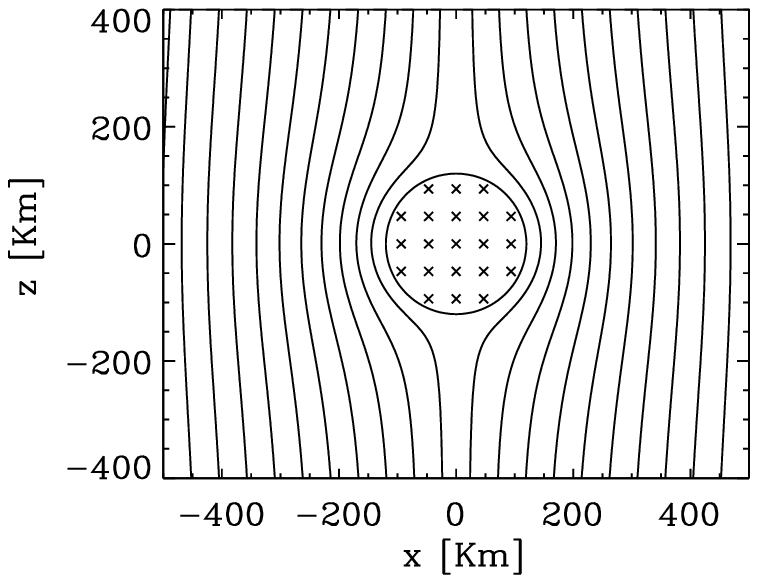}}
\resizebox{.39\hsize}{!}{\includegraphics[bb=0 5 283 188,clip]{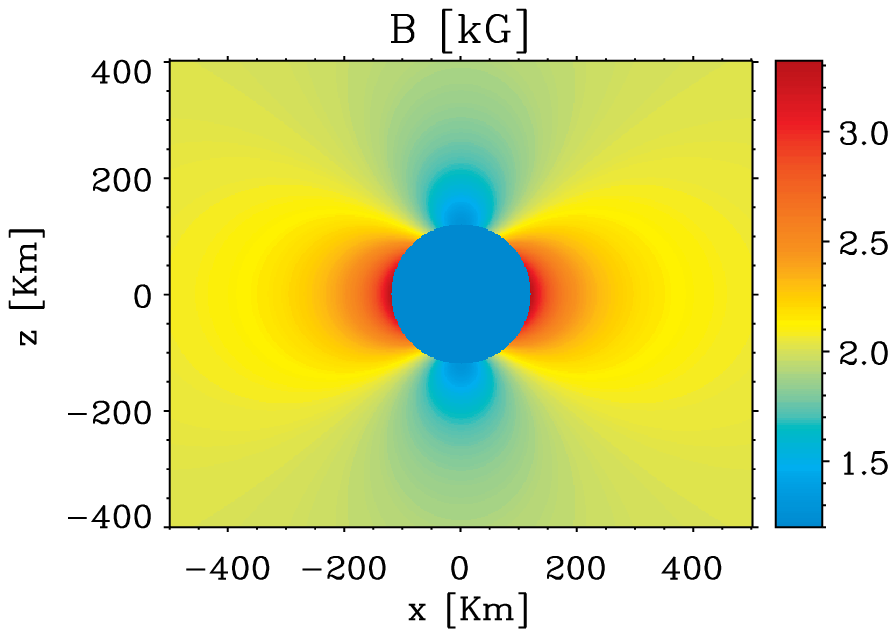}} \\
\resizebox{.39\hsize}{!}{\includegraphics[bb=0 0 283 203,clip]{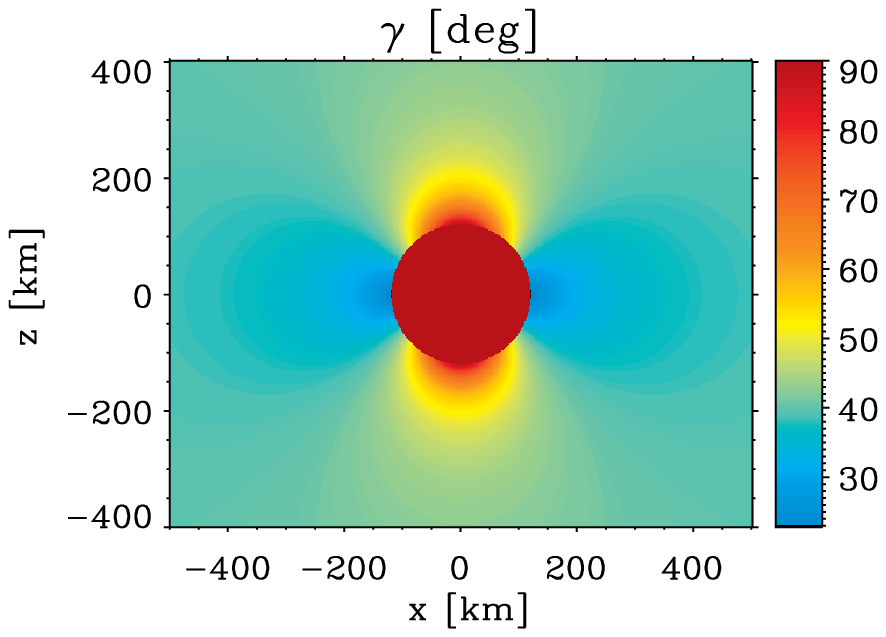}}
\resizebox{.39\hsize}{!}{\includegraphics[bb=0 0 283 203,clip]{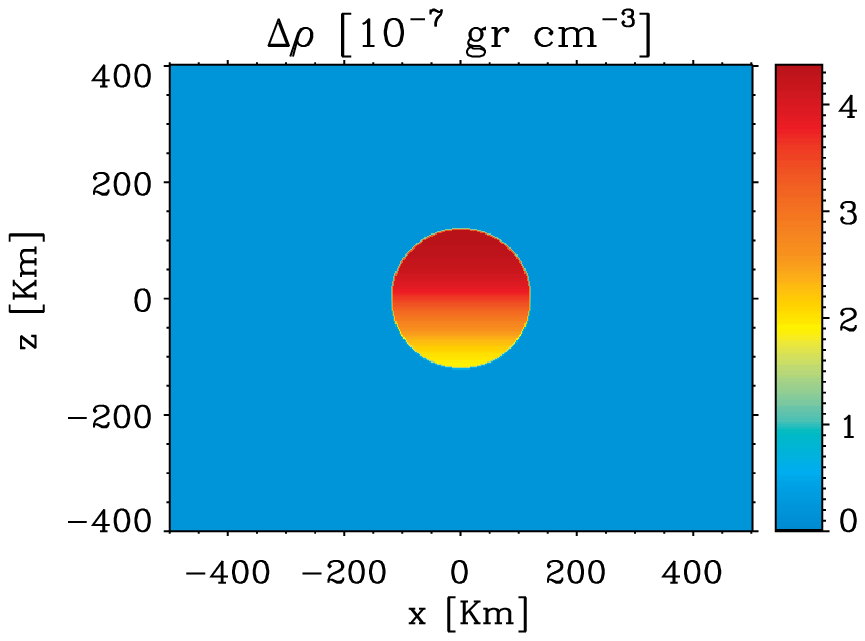}}
\end{center}
\vspace*{-1.3em}
\caption{{\em Top left:} magnetic field lines in the $xz$-plane. The
circle centered at $(0,0)$~km represents the flux tube's
boundary. Note the wrapping of the field lines around the tube. {\em
Top right:} field strength distribution.  {\em Bottom left:} field
inclination distribution.  {\em Bottom right:} gas density distribution, 
for temperatures in the tube and background given by the cool model of
Collados et al.\ (1994). Shown are density differences with respect to
the unperturbed atmosphere (Fig.~\ref{modeloinicial}).}
\label{magneticfield}
\end{figure*}

Recently, Heinemann et al.\ (2007) have presented first attempts to
simulate the penumbra in 3D. The parameters governing the calculations
are still far from those of the real sun and, as a consequence, the
model sunspot does not show a typical penumbral pattern. Yet, an
interesting result of the simulations is the existence of small blobs
of plasma with weaker and more inclined fields than their
surroundings. The magnetic properties of these structures are
reminiscent of those of the flux tubes postulated by the uncombed
model. Some of them show a dark lane similar to the dark cores of
penumbral filaments. The dark lanes are produced by locally enhanced
density and pressure that shift the $\tau=1$ level to higher
photospheric layers, where the temperature is lower. This effect was
identified for the first time by Sch\"ussler \& V\"ogler (2006) in
magnetoconvection simulations of umbral dots. Interestingly, the
parameter regime covered by those simulations is not the one relevant 
to sunspot penumbrae.

Our aim here is to shed some light on the origin of dark-cored
penumbral filaments and the surplus brightness of the penumbra. To
that end we solve the 2D stationary heat transfer equation in a
stratified uncombed penumbra formed by magnetic flux tubes in a
stronger background field (Sects.~\ref{model} and
\ref{equations}). The tubes carry an Evershed flow of hot plasma. Our
calculations show that one such tube would be observed as a
dark-cored filament due to the higher density of the plasma within the
tube (Sect.~\ref{results}). We also find that the Everhed flow heats
the background atmosphere very efficiently, increasing its
temperature. This suggests that the surplus brightness of the penumbra
is due to the Evershed flow (Sect.~\ref{surplus}). Finally, we
synthesize polarization maps using the model atmospheres resulting
from the simulations and compare them with polarimetric observations
of dark-cored filaments (Sect.~\ref{stokes_maps}).

\section{The model}
\label{model}
We describe a bright filament in the inner penumbra 
as a cylindrical flux tube of radius $R$ embedded in a stratified
background (umbral) atmosphere. The calculations are performed in 
a Cartesian coordinate system where the $z$-axis coincides with the
vertical and the $y$-direction is defined such that the axis of the tube
lies in the $yz$-plane. Since our primary goal is to identify the
mechanism(s) responsible for the existence of dark cores, we adopt the
simplest magnetic configuration possible, namely a potential
field. The field is determined from the conditions $\nabla
\cdot \vec{B} =0$ and $\nabla \times \vec{B} = 0$ in the $xz$-plane,
i.e., we neglect variations along the tube axis because they
are much smaller than variations perpendicular to it (e.g., 
Borrero et al.\ 2004).

In the tube's interior, the magnetic field is taken to be homogeneous
and directed along its axis, which is inclined by an angle
$\gamma_{\rm t}$ to the vertical: $\vec{B} = B_{\rm t} (\sin
\gamma_{\rm t} \vec{e}_{y} + \cos \gamma_{\rm t} \vec{e}_{z})$. Far
from the tube we assume the background magnetic field to be of the
form $\vec{B}= B_{\rm b} (\sin \gamma_{\rm b} \vec{e}_{y} + \cos
\gamma_{\rm b} \vec{e}_{z}$), so that the field is homogeneous with
strength $B_{\rm b}$ and inclination $\gamma_{\rm b}$. We use these
conditions, together with the continuity of the radial components at
the tube's boundary, to solve Laplace's equation in the
$xz$-plane. Spectropolarimetry tells us that the field 
is weaker and more inclined in the tubes (e.g., Bellot Rubio et al. 2004;
Borrero et al.\ 2004), hence we set $B_{\rm t} < B_{\rm b}$ and
$\gamma_{\rm t} > \gamma_{\rm b}$. The analytic expressions obtained
from the calculations are given in Appendix A.

Figure \ref{magneticfield} shows the magnetic configuration of the
model with $R= 120 $~km, $B_{\rm t}= 1200$~G, $\gamma_{\rm
t}=90^\circ$, $B_{\rm b}=2000$~G, and $\gamma_{\rm b}=40^\circ$. The
background field lines wrap around the tube ({\em top left}),
leading to enhanced field strengths on either side of the tube and
reduced field strengths above and below it ({\em top right}). 
The inclination of the background field also changes depending on
the position. In particular, the field above the tube is always more 
horizontal than $\gamma_{\rm b}$ due to the continuity of the normal
components at the tube's boundary ({\em bottom left}).

\begin{figure}
\begin{center}
\resizebox{1\hsize}{!}{\includegraphics[bb=17 25 386 271]{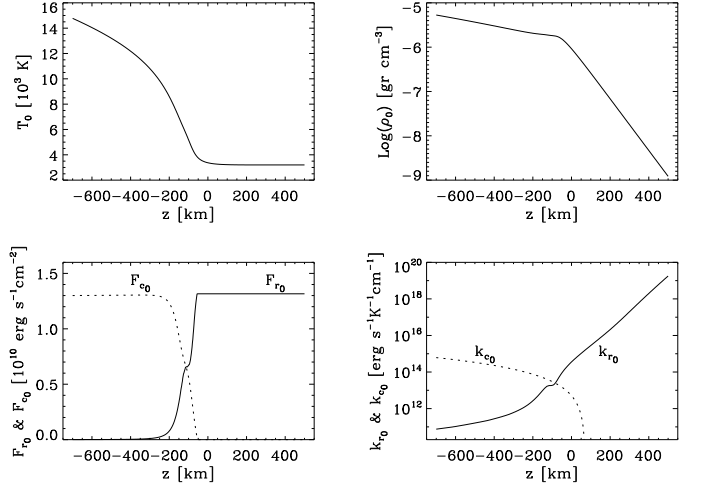}}
\end{center}
\caption{Physical parameters in the cool umbral model of Collados 
et al.\  
%\protect\linebreak[4] 
(1994). {\em Clockwise from upper left}: temperature, gas
density, thermal coefficients, and energy fluxes. Radiative and
convective parameters are shown in solid and dashed lines,
respectively.}
\label{modeloinicial}
\end{figure}

We require lateral mechanical equilibrium, therefore the total (gas
plus magnetic) pressure on both sides of the interface separating the
tube from the ambient medium is the same at the same geometrical
height $z$.  The magnetic field and the temperature then determine the
gas density through the ideal gas law (with variable mean molecular
weight to account for partial ionization). The equilibrium density
distribution shown in the bottom right panel of
Fig.~\ref{magneticfield} corresponds to the case in which the
temperatures of the tube and the external medium are those displayed
in Fig.~\ref{modeloinicial}. As can be seen, a strong density
enhancement occurs within the tube to compensate for its lower
magnetic pressure.

The condition of lateral force balance does not
guarantee vertical equilibrium, so the tube may stretch in the
vertical direction. Using simple estimates, however, Borrero et al.\
(2006) have shown that the vertical stretching could be limited by
buoyancy in the subadiabatic layers of the photosphere.
Alternatively, non-potential fields may ensure vertical force
equilibrium (Borrero 2007). Since there is no generally accepted
solution to this problem, we restrict ourselves to the simple magnetic
configuration described above in the hope that the results will not
depend significantly on the exact topology of the field.

\section{Heat transfer equation}
\label{equations}
To obtain the temperature distribution in the $xz$-plane we consider 
the stationary ($\partial/\partial t = 0$) heat transfer equation
\begin{equation}
\nabla \cdot \vec{F} = S,
\label{heattransfereq}
\end{equation}
neglecting gradients in the $y$-direction. Here, $\vec{F}=
\vec{F}_{\rm r} + \vec{F}_{\rm c}$ represents the heat flux vector and
$S$ the various energy sources, including Ohmic dissipation and the
Evershed flow. The radiative flux $\vec{F}_{\rm r}$ is computed using
the diffusion approximation
\begin{equation}
\vec{F}_{\rm r} = - \kappa_{\rm r} \, \nabla T
\label{flujoradiativo}
\end{equation}
(e.g., Mihalas 1978), with $\kappa_{\rm r}$ the radiative thermal
conductivity and $T$ the temperature.  Following Schlichenmaier et
al.\ (1999), we take $\kappa_{\rm r} = 16 \, D_{\rm F} \,
\tilde{\sigma} T^3/(k_{\rm R} \rho)$, where $\tilde{\sigma}$ is
Stefan-Boltzmann constant, $k_{\rm R}$ the Rosseland mean opacity,
$\rho$ the gas density, and $D_{\rm F}$ the flux limiter originally
introduced by Levermore \& Pomraning (1981). The convective flux
$\vec{F}_{\rm c}$ is evaluated using a linearized mixing length
approach (Moreno-Insertis et al.\ 2002),
\begin{equation}
\vec{F}_{\rm c}= -\kappa_{\rm c} \, [\nabla T - ({\rm d}T/{\rm d}z)_{\rm ad} ]
\equiv -\kappa_{\rm c} \, \nabla_{\rm c} T,
\label{flujoconvectivo}
\end{equation}
where $({\rm d}T/{\rm d}z)_{\rm ad} = T ({\rm d}P/{\rm d}z)
\nabla_{\rm ad}$ represents the adiabatic temperature gradient and
$\nabla_{\rm ad}$ the double-logarithmic isentropic temperature
gradient. The computation of $\nabla_{\rm ad}$ takes into account the
local physical conditions and the partial ionization of hydrogen (Cox
\& Giuli 1968). In our simulations,  $\nabla_{\rm ad}$ varies between 
0.1 and 0.4 depending on the layer. The convective transport
coefficient $\kappa_{\rm c}$ is evaluated following Spruit (1977). To
reduce the efficiency of convection in the presence of magnetic fields
we use $T_{\rm eff}=3888$~K, $z_1=70$~km, and $k_0=0.8 \times
10^{-11}$. The calculations are not very sensitive to the details of
the energy transport by convection because $\kappa_{\rm c}$ is
negligible in photospheric layers. We note, however, that a non-linear
treatment of the convective flux, with $\kappa_{\rm c}$ varying with
the changing superadiabaticity of the atmosphere, would be more
appropriate to deal with large temperature perturbations
(Sect.~\ref{evershed_energy}).

In the absence of penumbral flux tubes, the background atmosphere is a
plane-parallel stratified medium where $S=0$ and the physical
parameters do not vary with $x$ and $y$. Let us label them with the
subscript 0, i.e., $T_0(z)$, $\kappa_{\rm r0}(z)$, $\kappa_{\rm
c0}(z)$, and so on. Flux tubes embedded in this atmosphere present an
obstacle to the heat flow because of their higher gas density, which
decreases $\kappa_{\rm r}$. In addition, $S \neq 0$ within the
tubes. As a result, the temperature gets modified from $T_0(z)$ to
$T(x,z)$, but the gas pressure remains the same to ensure horizontal
force balance. To find the new equilibrium configuration we solve
Eq.~(1) iteratively. Let $T_i$ be the value of $T(x,z)$ in the
$i$-th step. This approximate solution does not satisfy Eq.~(1) by 
an amount
\begin{equation}
\epsilon_i \equiv S_i -\nabla \cdot [-{\kappa_{\rm r}}_i \nabla T_i 
- {\kappa_{\rm c}}_i \nabla_{\rm c} T_i] ,    \label{eq3}
\end{equation} 
where $S_i \equiv S(T_i)$, ${\kappa_{\rm r}}_i \equiv \kappa_{\rm r}(T_i)$, and
${\kappa_{\rm c}}_i \equiv \kappa_{\rm c}(T_i)$. Our goal is to find a
perturbation $\delta T_{i+1}$ such that the new temperature $T_{i+1}
\equiv T_i + \delta T_{i+1}$ leads to smaller errors, i.e., 
$\epsilon_{i+1} < \epsilon_{i}$.

In terms of $T_{i+1}$, the heat transfer equation can be written as 
\begin{equation}
    \nabla \cdot [-{\kappa_{\rm r}}_{i+1} \nabla T_{i+1} - 
{\kappa_{\rm c}}_{i+1}\nabla_{\rm c} T_{i+1}]= S_{i+1}.
    \label{eq5}
\end{equation} 
Defining
\begin{eqnarray}
   {\kappa_{\rm r}}_{i+1} = {\kappa_{\rm r}}_0 + \delta_0 {\kappa_{\rm r}}_{i+1} \nonumber,\\
   {\kappa_{\rm c}}_{i+1} = {\kappa_{\rm c}}_0 + \delta_0 {\kappa_{\rm c}}_{i+1} ,
   \label{eq4}
\end{eqnarray}
and neglecting the second-order terms $\delta_0 {\kappa_{\rm r}}_{i+1} 
\nabla (\delta T_{i+1})$ and $\delta_0 {\kappa_{\rm c}}_{i+1} 
{\nabla}_{\rm c} (\delta T_{i+1})$, Eq.~(\ref{eq5}) becomes
\begin{equation}
\nabla \cdot [-{\kappa_{\rm r}}_{0} \nabla \delta T_{i+1} - {\kappa_{\rm c}}_0 
\nabla_{\rm c} \delta T_{i+1}]= S_{i+1}-\nabla \cdot [-{\kappa_{\rm r}}_{i+1} 
\nabla T_{i} - {\kappa_{\rm c}}_{i+1} \nabla_{\rm c} T_{i}].
\label{eq6}
\end{equation}
The heat flux variation due to temperature variations is caused
primarily by changes in the temperature gradient, while the effects of
changes in the transport coefficients are negligible to first order
(Moreno-Insertis et al.\ 2002). This implies
\begin{eqnarray}
{\kappa_{\rm r}}_{0} \nabla \delta T_{i+1}  & >> & [{\kappa_{\rm r}}_{i+1}-{\kappa_{\rm r}}_{i}]\nabla T_{i}, \\
{\kappa_{\rm c}}_{0} \nabla_{\rm c} \delta T_{i+1} & >> & [{\kappa_{\rm c}}_{i+1}-{\kappa_{\rm c}}_{i}]\nabla_{\rm c} T_{i},
\label{eq7}
\end{eqnarray} 
allowing us to make the approximations ${\kappa_{\rm r}}_{i+1} \simeq {\kappa_{\rm r}}_{i}$
and ${\kappa_{\rm c}}_{i+1} \simeq {\kappa_{\rm c}}_{i}$ in Eq.~(\ref{eq6}). With the
additional assumption that $S_{i+1} \simeq S_{i}$, Eq.~(\ref{eq6}) can finally be 
rewritten as
\begin{equation}
\nabla \cdot [-{\kappa_{\rm r}}_{0} \nabla \delta T_{i+1} - {\kappa_{\rm c}}_0 
\nabla_{\rm c} \delta T_{i+1}]= \epsilon_i,
\label{eq8}
\end{equation}
where the coefficients on the left-hand side depend only on $z$. We
seek for periodic solutions of Eq.~(\ref{eq8}) in the
$x$-direction. Thus, the following Neumann boundary conditions are
imposed:
\begin{eqnarray}
\delta T_{i}(x,z=z_{\rm min})  = 0   &\qquad&   \forall x, \\
\dot{\delta T}_i(x=0,z) =\dot{\delta T}_i(x=x_{\rm max},z) = 0   &\qquad&  \forall z, 
\end{eqnarray} 
where the dots represent derivation with respect to $x$. 
The first condition prescribes zero temperature variations at the
bottom of the computational domain, while the second forces the
temperature perturbations to be minimum at the lateral boundaries. 
With these ingredients, Eq.~(\ref{eq8}) is solved numerically
for each height $z$ using Fourier cosine transforms in the
$x$-direction. Convergence ($\mbox{max} \, |{\epsilon_i/S_i}|<10^{-4}$) 
is achieved in 5 to 10 iterations.

\subsection{Simulation setup}

\begin{figure}
\begin{center}
%\resizebox{.99\hsize}{!}{\includegraphics[bb=8 15 272 120]{tubitoat11.eps}}
\resizebox{1\hsize}{!}{\includegraphics[bb=7 19 328 138]{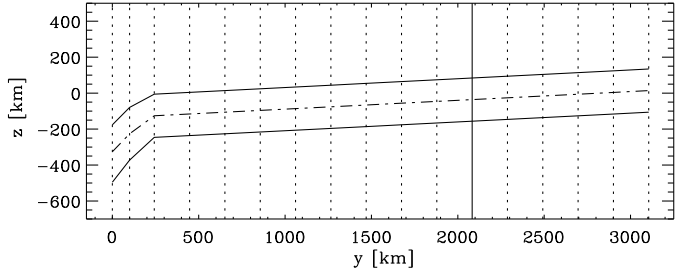}}
\end{center}
\caption{Cut of the computational domain at $x=0$ km, showing a flux
tube 120 km in radius. The vertical dotted lines mark the position of 
the 17 $xz$-planes where we solve the stationary heat transfer equation. 
The vertical solid line at $y= 2083$~km indicates the position of the 
$xz$-cut displayed in Figs.\ \ref{sinevershed} and \ref{conevershed}.}
\label{tubito}
\end{figure}

The computational domain is a cube extending 1000 km in the
$x$-direction, 3100 km in the $y$-direction, and 1200 km in the
$z$-direction. Initially, the cube is filled with an unperturbed
umbral atmosphere (the cool model of Collados et al.\ 1994) with field
strength $B_{\rm b}=2000$~G and inclination $\gamma_{\rm b} =
40^\circ$.  Figure 2 shows the run with depth of the temperature, gas
density, and radiative and convective parameters in the unperturbed
model. A magnetic flux tube is inserted in this atmosphere. It 
has a radius $R= 120$ km, a current sheet 2 km thick, a field strength
$B_{\rm t}= 1200$~G, and field inclinations varying between 45$^\circ$
and 87$^\circ$ as indicated in Fig.~\ref{tubito}. The axis of the tube
is placed at a height determined by the field inclination and the
distance along the tube, starting with $z=-326$~km at $y=0$~km.

The heat transfer equation is solved in 17 $xz$-planes at different
$y$-values (Fig.~\ref{tubito}). Each
plane is discretized in $501 \times 601$ grid points separated by 2
km.

\section{Results}
\label{results}
We first consider Ohmic dissipation as the only source of energy
(Sect.~\ref{ohmic}). Thereafter, we examine how the results are modified 
by a hot Evershed flow along the magnetic flux tube
(Sect.~\ref{evershed_energy}).

\subsection{Ohmic dissipation as the only source of energy}
\label{ohmic}
In this case, the source term of the heat transfer equation reduces to
$S = j^2/\sigma$, where $j$ is the electric current associated with
the jump of $\vec{B}$ at the tube's boundary (Eq.~\ref{corrientes}) 
and $\sigma$ the electric conductivity, computed following Kopeck\'y 
\& Kuklin (1969).

\begin{figure}
\begin{center}
\resizebox{1\hsize}{!}{\includegraphics[bb= 0 230 386 664]{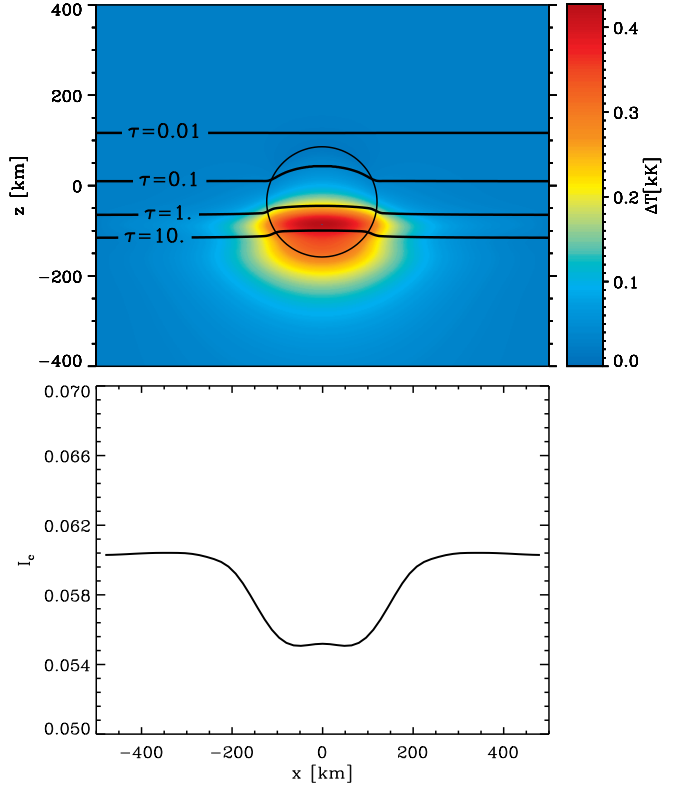}}
\end{center}
\caption{{\em Top:} temperature perturbations $\Delta T \equiv
T(x,z)-T_0(z)$ in the \protect\linebreak[4] $xz$-plane at $y =2083$
km. Ohmic dissipation is the only energy source considered. The circle
represents the flux tube. Solid lines are lines of constant Rosseland
optical depth. {\em Bottom:} continuum intensity at 487.8~nm emerging
from the $y= 2083$ km plane, convolved with the Airy point-spread
function of a 1-m telescope. The values are normalized to the quiet
Sun continuum intensity $I_{\rm QS}$.}
\label{sinevershed}
\end{figure}

Figure~\ref{sinevershed} shows the simulation results at
$y=2083$~km. The tube blocks the energy coming from below and this
produces a strong heating of its lower half and the adjacent medium. The
associated temperature enhancements, however, are difficult to detect
because they occur below the photosphere for the most part.

The continuum intensity observed at the surface is determined mainly
by the temperature prevailing at Rosseland optical depth unity,
$\tau_{\rm R}=1$. In the top panel of Fig.~\ref{sinevershed}, lines of
constant optical depth are indicated.  As can be seen, the $\tau_{\rm
R}=1$ level is shifted upwards within the flux tube. The reason is the
larger gas density in the tube compared with the external medium,
which results from its weaker field strength and the condition of
horizontal mechanical equilibrium. The enhanced density increases the
opacity and moves the $\tau_{\rm R}=1$ level to higher layers, where
the temperature is lower owing to the stratification of the model. In
accordance with the Eddington-Barbier relation, the continuum
intensity decreases at the position of the tube.

The bottom panel of Fig.~\ref{sinevershed} shows the continuum
intensity emerging from the model at 487.8~nm, as computed by solving
the radiative transfer equation and convolving the result with the
theoretical Airy point-spread function of a 1-m telescope. The flux 
tube is darker than its surroundings and do not exhibit bright edges, 
in clear contradiction with the observations. To reproduce the
properties of dark-cored filaments, the walls of the tube should be
hotter than the background atmosphere. The $\tau_{\rm R} =1$ level
would then encounter higher temperatures at the tube's boundary,
generating two lateral brightenings. Ohmic dissipation in the current
sheet, however, cannot provide the necessary heating because of the
high conductivity of the plasma. In fact, one would have to reduce the
conductivity by at least two orders of magnitude, or to consider
extremely thin current sheets (a few meters thick), to heat the walls
of the tube to the required degree. Even in that case the dark core
would still show smaller continuum intensities than the surroundings.
It is therefore necessary to heat not only the walls of the tube, but
also the tube itself. In this context, the Evershed flow represents a
natural source of energy.

\subsection{Heating by the Evershed flow}
\label{evershed_energy}

\begin{figure}
\begin{center}
\resizebox{1\hsize}{!}{\includegraphics[bb= 0 230 386 664]{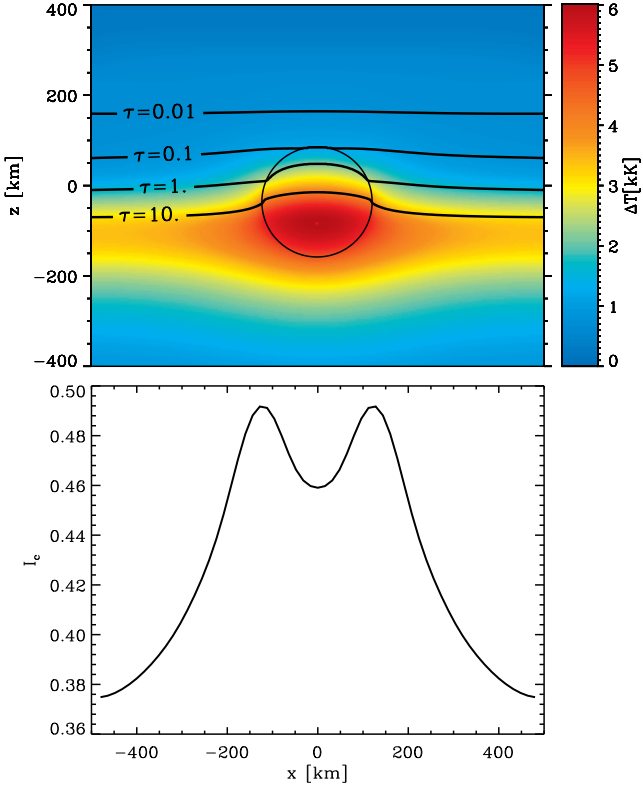}}
\end{center}
\caption{Same as Fig.~\ref{sinevershed}, for the case in which the
sources of energy are Ohmic dissipation and a hot Evershed flow. }
\label{conevershed}
\end{figure}

The Evershed flow is a radial outflow of mass associated with the more
inclined magnetic fields of the penumbra (Title et al.\ 1993;
Stanchfield et al.\ 1997; Schlichenmaier \& Schmidt 2000; Westendorp
Plaza et al.\ 2001; Borrero et al.\ 2005; Bellot Rubio et al.\ 2006;
Rimmele \& Marino 2006; Ichimoto et al.\ 2007).

An Evershed flow of hot plasma along a magnetic flux tube produces an
energy flux $\vec{F}_{\rm E}$ whose divergence can be evaluated
from the entropy equation as
\begin{equation}
\nabla \cdot \vec{F}_{\rm E} = \rho c_V \vec{v}_{\rm E} \, [\nabla T - ({\rm
d}T/{\rm d}z)_{\rm ad} \vec{e}_{\rm z}], 
\label{flujoevershed}
\end{equation}
where $c_V$ is the specific heat at constant volume and $\vec{v}_{\rm
E}$ the flow velocity ($\vec{v}_{\rm E} =0$ outside the tube). To
account for the heating induced by the Evershed flow, we set
$S=j^2/\sigma + \nabla \cdot \vec{F}_{\rm E}$ in Eq.~(\ref{heattransfereq}). 
We assume that the flow is parallel to the magnetic field  
(Bellot Rubio et al.\ 2004), and that its velocity changes 
with radial distance as dictated by mass conservation due to the 
density decrease toward higher layers.

The calculation of Eq.~(\ref{flujoevershed}) cannot be done in 2D
because of the term $\nabla T$. To solve the problem we follow an
iterative approach. $\nabla T$ is estimated from the temperatures in
the plane $y=y_{j-1}$, $T(x,y_{j-1},z)$, and a guess for the
temperatures at $y=y_j$, $T_i(x,y_{j},z)$. Once $S$ is known, the
method described in Sect.~\ref{equations} can be applied to the
$y=y_j$ plane. This results in a new temperature distribution $T_{i+1}
(x,y_{j},z)$ which is used to update $\nabla T$ for the next
iteration. Convergence is reached in 10--50 steps. To initialize 
the calculations, the temperature distribution in the
$y= y_j$ plane is taken to be the temperature at $y= y_{\rm j-1}$,
plus the temperature difference between the two planes in the
non-Evershed case. 

The temperature excess induced by the flow at $y=0$~km is prescribed
as a boundary condition. This condition determines the overall
brightness of the flux tube and its surroundings, so it is relatively
well constrained. In our calculations we use a temperature excess of
7500~K within the tube at $y=0$~km, to ensure that its brightness is
compatible with the observations. An excess of 7500~K is also
compatible with the maximum temperatures of $\sim$13\,500~K that hot
plasma rising adiabatically from the bottom of the convection zone
would show at photospheric levels (Borrero 2007).

The upper panel of Fig.~\ref{conevershed} displays the simulation
results at $y=2083$~km, assuming a flow velocity of $v_{\rm E}=
7$~km~s$^{-1}$ in the first plane.  The equilibrium temperature
distribution is characterized by an intense heating of the tube and
the surroundings, with temperature enhancements of up to 6000~K. This
heating increases the brightness of the external medium and modifies
the opacity of the plasma in a way that produces filaments with a dark
core and two lateral brightenings (lower panel of
Fig.~\ref{conevershed}).

\begin{figure}
\begin{center}
%\resizebox{1\hsize}{!}{\includegraphics[bb=15 20 285 163]{fig5at8.eps}}
%\resizebox{1\hsize}{!}{\includegraphics[bb=10 20 280 163]{fig5at11_2008.eps}}
%\resizebox{1\hsize}{!}{\includegraphics[bb=10 20 280 163]{fig5at11bis2.eps}}
%\resizebox{1\hsize}{!}{\includegraphics[bb=10 20 280 170]{fig5a_new.eps}}
\resizebox{1\hsize}{!}{\includegraphics[bb=10 10 280 171]{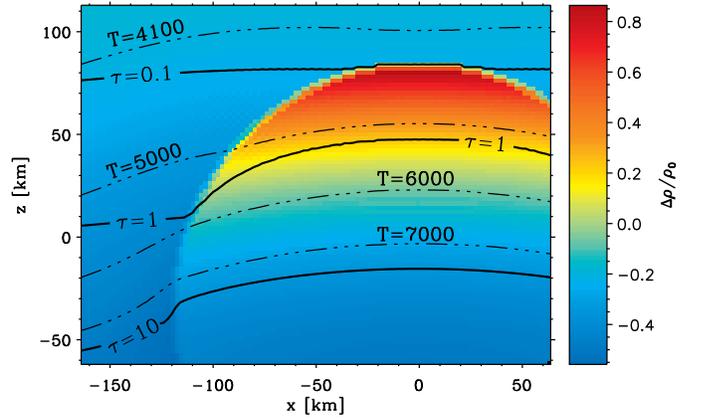}}
\end{center}
\caption{Section of the $xz$-plane at $y=2083$~km showing gas density
perturbations $\Delta \rho/\rho_0 = [\rho (x,z) -
\rho_0(z)]/\rho_0(z)$ induced by the weaker field of the tube and the
Evershed flow. Dash-dotted lines represent isotherms.  Solid lines
indicate lines of constant Rosseland optical depth. } 
\label{explicacion}
\end{figure}

Figure~\ref{explicacion} examines in greater detail the origin of
these structures. We show the gas density perturbations, $\Delta \rho
/\rho_0 = [\rho(x,z) - \rho_0(z)]/\rho_0(z)$, in the $xz$-plane at
$y=2083$~km together with lines of constant temperature ({\em
dash-dotted}) and constant Rosseland optical depth ({\em solid}). The
heating caused by the Evershed flow shifts the $\tau_{R}$ lines to
higher layers in and near the tube. For $\tau_{\rm R} \geq 10$, the
lines of constant $\tau_{\rm R}$ follow the isotherms closely, i.e.,
deep in the atmosphere the optical depth scale is primarily determined
by the temperature, not by the gas density. This is due to the strong
temperature dependence of the H$^{-}$ opacity. As the tube is
approached from $x= -150$ km, the $\tau_{\rm R}=1$ level crosses
isotherms corresponding to higher temperatures, which explains the
larger continuum intensities of the filament compared with its
surroundings. Very close to the tube's boundary the $\tau_{\rm R}=1$
line encounters regions of even higher temperatures, producing a
maximum in the continuum intensity (the lateral brightenings). As soon
as the $\tau_{\rm R} =1$ line enters the tube it evolves in a region
of enhanced gas density. The increased opacity moves the $\tau_{\rm
R}=1$ surface upward by about 40~km, where the temperature is
cooler. This generates a dark core.

The density enhancement is more prominent near the top of the tube
because the difference of magnetic pressures between the interior and
the external medium attains a maximum there with respect to the
external gas pressure. The temperature perturbations induced by the
Evershed flow decrease the large density contrasts resulting from the
weaker field of the tube (bottom right panel of
Fig.~\ref{magneticfield}), but this effect is less important in
determining the equilibrium density distribution.

Two interesting facts deserve further consideration. First, the
$\tau_{\rm R}=0.1$ level is reached at the top of the tube, some 40~km
above $\tau_{\rm R}=1$. Thus, a substantial fraction of the line
formation region is contained within the flow channel. Second, the density
distribution depicted in Fig.~\ref{explicacion} tends to stabilize the
tube against vertical stretching, since it causes negative buoyancy at
the top of the tube. A similar density distribution has been found by
Borrero (2007) under the assumption of flux tubes in exact force
balance.

\begin{figure}
\begin{center}
%\resizebox{1\hsize}{!}{\includegraphics[bb=13 23 335 103]{figlast_6p3_new.eps}}
%\resizebox{1\hsize}{!}{\includegraphics[bb=10 23 437 103]{figlast_6p3_new.eps}}
%\resizebox{1\hsize}{!}{\includegraphics[bb=10 23 437 103]{figlast003_6p3.eps}}
%\resizebox{1\hsize}{!}{\includegraphics[bb=10 23 435 115]{figlast_1104.eps}}
\resizebox{1\hsize}{!}{\includegraphics[bb=10 23 435 115]{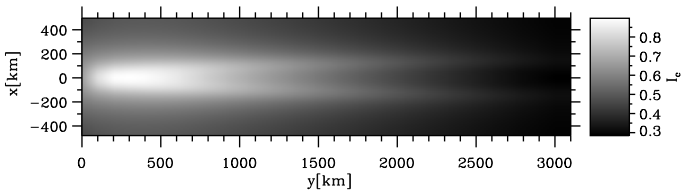}}
%\resizebox{1\hsize}{!}{\includegraphics[bb=2 15 275 70]{fig6_p3a.eps}}
\end{center}
\caption{Dark-cored penumbral filament produced by the flux tube
considered in this work as seen in continuum intensity at 487.8~nm.
The Evershed flow has a velocity of $v_{\rm E} = 7$~km~s$^{-1}$ and
induces a temperature excess of $7500$~K within the tube at $y=0$~km.}
\label{intensidad2D}
\end{figure}

Figure~\ref{intensidad2D} displays a continuum (487.8~nm) image of the
tube as it would appear through a 1-m telescope at disk
center. Clearly, the morphological properties of dark-cored filaments
are well reproduced: the inner footpoint shows up as a bright
penumbral grain, the central dark lane is surrounded by two lateral
brightenings separated by roughly 250~km ($\sim$0\farcs3), the
intensity ratio between the dark core and the lateral brightenings is
0.93, and the dark core can be traced for more than 2000~km (Scharmer
et al.\ 2002; S\"utterlin et al.\ 2004; Rouppe van der Voort et al.\
2004; Bellot Rubio et al.\ 2005; Rimmele 2008). We also note that the
distance between the lateral brightenings reflects the true diameter
of the underlying magnetic flux tube, at least at the resolution of a
1-m telescope.

\section{Discussion}
\label{discussion}

\subsection{Origin of dark-cored penumbral filaments}

Schlichenmaier et al.\ (1999) studied the radiative cooling of hot
flux tubes surrounded by an initially isothermal atmosphere. They
found that the tubes cool down quickly in the absence of energy
sources, reaching thermal equilibrium with the external medium in only
a few tens of seconds.  In the case of optically thick tubes (initial
temperatures above 10\,000~K), a steep cooling front develops and
migrates towards the tube axis at constant velocity, while optically
thin tubes (initial temperatures below 7500~K) cool more or less
homogeneously over their cross sections. In either case, no dark cores
would be observed.

Our simulations can be regarded as an extension of the work by
Schlichenmaier et al.\ (1999). The main difference is that we consider
a stratified atmosphere consisting of magnetic flux tubes in a
stronger ambient field. Under these conditions, the equilibrium
temperature distribution is no longer symmetric around the tube
axis. We have demonstrated that one such tube would exhibit a central
dark core due to the higher density (hence larger opacity) of the
plasma within the tube, which moves the $\tau_{\rm R}=1$ level to
higher (cooler) layers. The same mechanism produces dark lanes in
umbral dots (Sch\"ussler \& V\"ogler 2006) and gappy penumbrae (Spruit
\& Scharmer 2006), although in these structures the density
enhancement is a consequence of overturning convection.  In an
uncombed penumbra, the density increase is essentially due to the
weaker field of the tubes.

It is important to remark, however, that the observations cannot be
fully explained without a hot Evershed flow along the tubes. The flow
increases the brightness of the flux tube relative to the surroundings
and generates the two lateral brightenings of the filaments; in the
absence of this energy source, the tubes would actually be darker than
their environs.

Borrero (2007) has also been able to reproduce dark-cored filaments
using thick penumbral flux tubes in magnetohydrostatic
equilibrium. The condition of vertical and horizontal force balance
determines the temperature distribution in and around the tubes, given
the magnetic field distribution. This results in structures that are
hotter than the external medium at the same geometrical height, in
good qualitative agreement with our calculations. Borrero (2007) did
not identify the source of the heating, but we hypothesize that the
higher temperatures required to maintain the tubes in force 
equilibrium are actually produced by the Evershed flow. To verify 
this conjecture, the energy and momentum equations must be solved
simultaneously in the presence of hot upflows.

\subsection{Surplus brightness of the penumbra}
\label{surplus}

An important result of the calculations described in
Sect.~\ref{evershed_energy} is that the background atmosphere 
itself is much brighter than it would be in the absence of an 
Evershed flow. The continuum intensity without flows is only 
0.06\,$I_{\rm QS}$ (Fig.~\ref{sinevershed}), corresponding to a 
very cool umbra. In contrast, when a hot upflow is considered, 
the background shows intensities of up to $\sim$0.7\,$I_{\rm QS}$ 
near the tube (Fig.~\ref{intensidad2D}). This enormous difference 
can explain the surplus brightness of the penumbra relative  
to the umbra. 

The average continuum intensity emerging from the box of
Fig.~\ref{intensidad2D} is $\sim$0.5\,$I_{\rm QS}$, i.e., slightly
smaller than the observed penumbral brightness but of the same order
of magnitude. Adjustments of the temperature in the background
atmosphere, the geometry of the flux tubes, and/or the boundary
conditions for the Evershed flow should easily lead to a closer match.

\subsection{Polarimetric signatures of dark-cored penumbral filaments}
\label{stokes_maps}

In this section we investigate the appearance of penumbral flux tubes
in polarized light using the results of
Sect.~\ref{evershed_energy}. We have chosen the \ion{Fe}{i} line at
630.25~nm to solve the radiative transfer equation because many
instruments, including the SOUP magnetograph at the Swedish Solar
Telescope (SST) and the polarimeters of the Solar Optical Telescope aboard
Hinode, have measured the properties of dark-cored filaments in this
line.

\begin{figure}
\begin{center}
\resizebox{.99\hsize}{!}{\includegraphics[bb=3 15 275 75,clip]{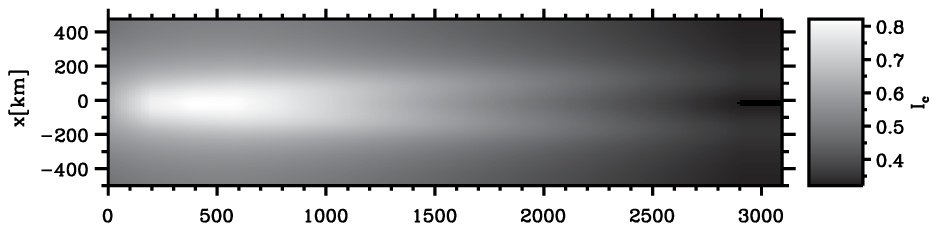}}
\resizebox{.99\hsize}{!}{\includegraphics[bb=3 15 275 75,clip]{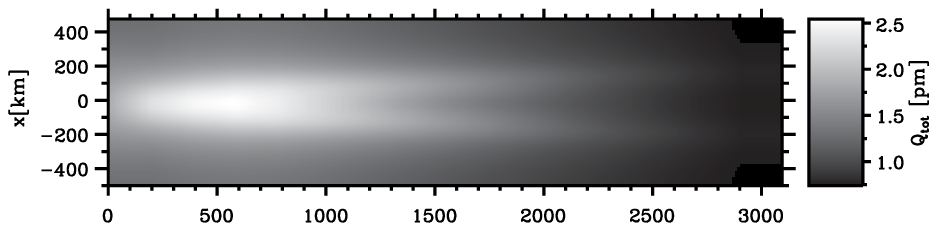}}
\resizebox{.99\hsize}{!}{\includegraphics[bb=3 15 275 75,clip]{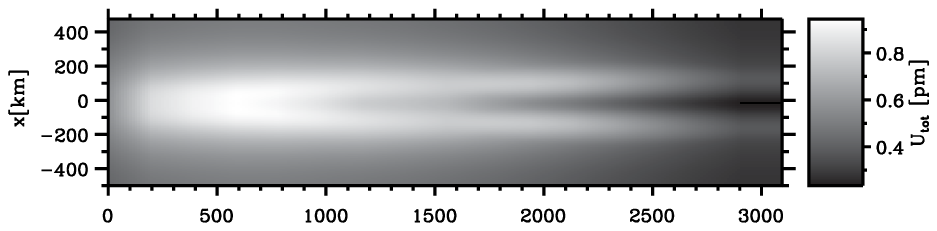}}
\resizebox{.99\hsize}{!}{\includegraphics[bb=3 0 275 75,clip]{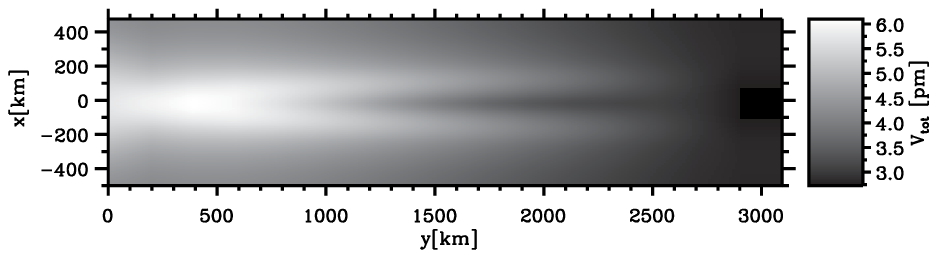}}
\resizebox{.99\hsize}{!}{\includegraphics[bb=3 15 275 80,clip]{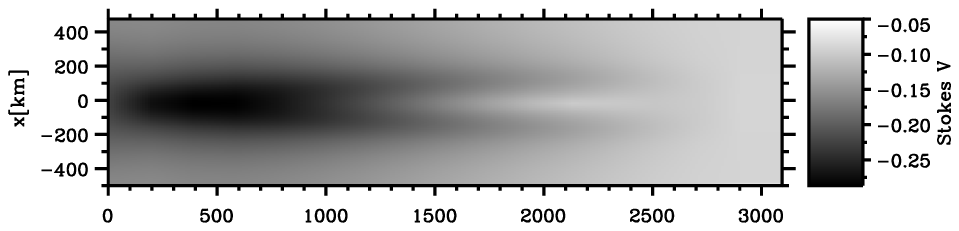}}
\resizebox{.99\hsize}{!}{\includegraphics[bb=3 3 275 75,clip]{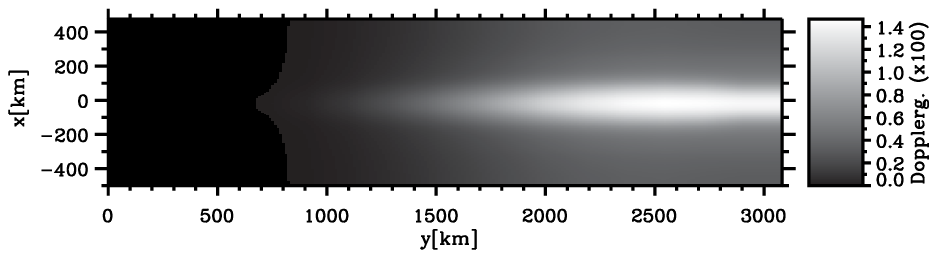}}
\end{center}
\caption{{\em Top:} polarization maps of the dark-cored penumbral 
filament resulting from the simulation of Sect.\ 4.2 in the
\ion{Fe}{i} 630.25~nm line. The four panels show the continuum intensity
and the wavelength-integrated (unsigned) $Q$, $U$, and $V$ polarization
through a 1-m telescope at disk center. The polarization profiles have been
normalized to the quiet Sun continuum. {\em Bottom:}
magnetogram at $+10$ pm from line center and Dopplergram at $\pm 15$
pm from line center. }
\label{stokes}
\end{figure}

The four top panels of Fig.~\ref{stokes} show the flux tube as it
would be recorded through a 1-m telescope at disk center in continuum
intensity and wavelength-integrated Stokes $Q$, $U$, and $V$
polarization, respectively. These maps can be compared with real
observations. The dark lane is more prominent in polarized light ($Q$,
$U$, and $V$) than in continuum intensity, as observed with Hinode
(Bellot Rubio et al.\ 2007) and the SST (van Noort \& Rouppe van
der Voort 2008). In total circular polarization, however,
the lateral brightenings disappear at too short a distance from the
filament head. We believe this is due to the geometry assumed for the
flux tube, which never becomes horizontal. Because of the heights
attained by the tube, the radiative cooling of the Evershed flow
proceeds at a very fast pace. Also, the tube leaves the region of
maximum sensitivity of Stokes $V$ comparatively soon, which decreases
the circular polarization signal. For these reasons, we expect that a
better choice of the tube inclination at large radial distances will
produce longer filaments in circular polarization.

The bottom panels of Fig.~\ref{stokes} display a synthetic magnetogram
and a Dopplergram of the filament. They have been constructed using
the Stokes $V$ profile of \ion{Fe}{i} 630.25~nm at $+10$ pm from line
center, and the corresponding Stokes $I$ profile at $\pm 15$ pm,
respectively. The magnetogram signal is weaker in the dark core than
in the lateral brightenings (white means less signal), suggesting
weaker and/or more inclined fields. This is exactly what has been
inferred from polarimetric measurements of dark-cored filaments
(Langhans et al.\ 2007; Bellot Rubio et al.\ 2007; van Noort \& Rouppe
van der Voort 2008). The Dopplergram shows a strong signal in the dark
core, indicating the existence of a flow there. Given the disk-center
position of the tube and the nearly horizontal inclination of the
flow, the Doppler shift is not very conspicuous in this particular
example; observations closer to the limb would certainly show stronger
signals. In any case, the Dopplergram of Fig.~\ref{stokes} provides
the same information as the high-resolution spectroscopic measurements
of Bellot Rubio et al.\ (2005), Rimmele \& Marino (2006), and Langhans
et al.\ (2007): the flow appears preferentially in the dark core, not
in the lateral brightenings or the adjacent medium.

The qualitative agreement between synthetic and observed parameters is
not surprising. As mentioned in Sect.~\ref{evershed_energy}, a
significant fraction of the line-forming region is contained within
the flux tube. This means that the physical conditions of the tubes
leave their signatures in the Stokes spectra of magnetically sensitive
lines. Since the tubes have weaker and more inclined fields, together
with strong flows, the polarization profiles emerging from them cannot
indicate otherwise. To some extent these arguments justify the
conclusions drawn from simple interpretations of polarimetric
measurements, but we caution that precise determinations of the
magnetic field vector and flow velocity of dark-cored penumbral
filaments will require sophisticated inversion techniques such as
those used by Borrero et al.\ (2005) or Jur\v{c}ak et al.\ (2007).

\section{Conclusions}

The heat transfer and radiative transfer calculations presented in
this paper support the concept of a penumbra formed by small (but
optically thick) magnetic flux tubes that carry hot flows, as deduced
from high-resolution observations and spectropolarimetric measurements
(see Solanki 2003 and Bellot Rubio 2004 for reviews). Tubes about
250~km in diameter explain not only the existence of dark-cored
filaments, but also the surplus brightness of the penumbra; the
Evershed flow efficiently heats the plasma outside the tubes,
increasing its temperature to values compatible with the observations.

Further improvements of the model should include a more realistic
treatment of the magnetic topology of the tubes and the external
atmosphere, a better description of convection, and a full 3D solution
of the heat transfer equation. In our opinion, however, these
improvements will not change the conclusion that the uncombed model 
is the best representation of the penumbra at our disposal.

%provides the best description of the structure of sunspot penumbrae 
%currently available.  
%is the one providing the most accurate description of the penumbra.
%most accurate description of the penumbra is still provided by the
%uncombed model. 

\begin{acknowledgements}
  Discussions with J.M.\ Borrero, F.\ Moreno-Insertis, and F.~P\'erez
  are gratefully acknowledged. We also thank the anonymous referee for
  his/her suggestion to check the behavior of the entropy, which
  allowed us to correct an error in the calculations of Sect.\
  4.2. This work has been supported by the Spanish Ministerio de
  Educaci\'on y Ciencia under projects AYA2007-63881,
  ESP2006-13030-C06-02, and PCI2006-A7-0624.
\end{acknowledgements}

\begin{appendix}
\section{Magnetic field distribution}

\begin{figure}
\begin{center}
\resizebox{.9\hsize}{!}{\includegraphics[bb = 0 0 257 304]{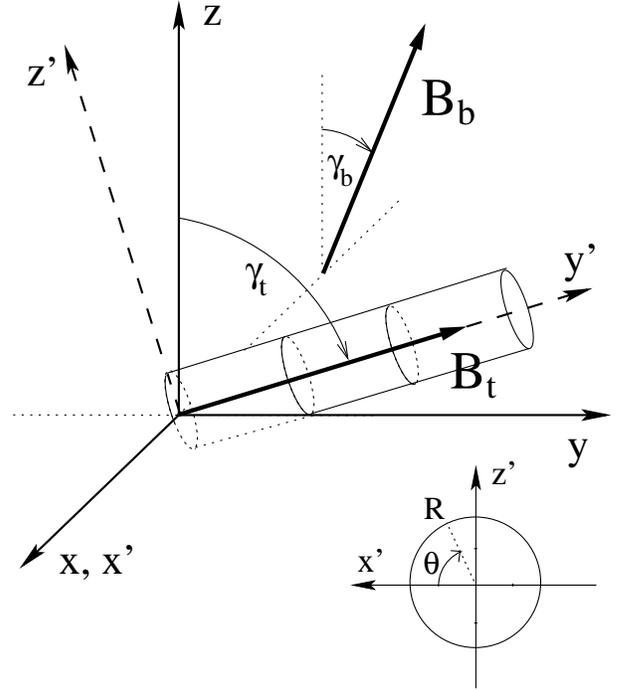}}
\end{center}
\caption{Coordinate systems used in the calculations. The $z$-axis represents
the normal to the solar surface. The tube axis is contained in the $yz$-plane.
$\gamma_{\rm t}$ and $\gamma_{\rm b}$ are the inclinations of the magnetic
field vector in the flux tube ($\vec{B_{\rm t}}$) and the background
atmosphere far from the tube ($\vec{B_{\rm b}}$), respectively. A new
reference system $x'y'z'$ is defined by rotating the $xyz$-system around the
$x$-axis until the new $y'$-axis coincides with the tube axis. The lower inset
shows the cross section of the tube and the angle $\theta$ of the cylindrical
coordinate system. }
\label{referencesystem}
\end{figure}

We take advantage of the information provided by observations and
numerical simulations to constrain the magnetic properties of the
model. As mentioned in Sect.\ 2, the field strength inside the tube,
$B_{\rm t}$, must be smaller than in the external medium, $B_{\rm
b}$. Also, the inclination of the field in the tube, $\gamma_{\rm t}$,
should be larger than in the background far from it, $\gamma_{\rm
b}$. These inclinations are measured with respect to the local
vertical, as sketched in Fig.~\ref{referencesystem}.

A potential field satisfies the conditions $\nabla \cdot \vec{B} =0$
and $\nabla \times \vec{B} = 0$. We solve this problem in cylindrical
coordinates $r\theta y'$, where the $y'$-axis coincides with the axis
of the tube (Fig.\ \ref{referencesystem}).
%Once the magnetic field
%vector is known in cylindrical coordinates, the transformation to the
%$xyz$ system is straightforward. 
In the tube's interior ($r<R-\delta$), the field is assumed to be 
homogeneous and directed along its axis, i.e.,
\begin{eqnarray}
B_r & = & B_\theta = 0, \nonumber \\
B_{y'} & = & B_{\rm t}.
\end{eqnarray}
Variations of the field along the $y'$-axis are ignored because
%, according 
%to spectropolarimetric analyses, 
they are much smaller than variations perpendicular to the tube's axis
(Borrero et al.\ 2004). In the background atmosphere, we assume that
far from the tube the magnetic field vector is inclined by an angle
$\gamma'_{\rm b} = 90 + \gamma_{\rm b} - \gamma_{\rm t}$ with respect
to the $z'$-axis, and that the field does not depend on $z'$.
Further, we require continuity of the radial component of the field at
$r=R$.  Using these boundary conditions, the potential solution for
the magnetic field in the external medium ($r > R$) reads
\begin{eqnarray}
B_r      &=& B_{\rm b} \, \cos \gamma'_{\rm b} \, \sin\theta \, (1-R^2/r^2), \nonumber \\
B_\theta &=& B_{\rm b} \, \cos \gamma'_{\rm b} \, \cos\theta \, (1+R^2/r^2), \\ 	
B_{y'}   &=& B_{\rm b} \, \sin \gamma'_{\rm b} \nonumber
\end{eqnarray}
Finally, in the current sheet ($r \in [R-\delta, R]$) we demand 
$B_\theta$ and $B_{y'}$ to vary linearly, so that
\begin{eqnarray}
B_r &= &  0, \nonumber \\
B_\theta &=& 2 \, B_{\rm b} \, \cos \gamma'_{\rm b} \, \cos \theta \, 
             \left[ r-(R-\delta) \right] /\delta, \\
B_{y'} &= & B_{\rm t} + (B_{\rm b} \, \sin 
                \gamma'_{\rm b}-B_{\rm t}) \left[(r-(R-\delta)\right]/\delta. \nonumber
\end{eqnarray}
With this choice, the magnetic field vector is continuous across the walls of
the tube and verifies the condition $\nabla \cdot \vec{B}=0$.

The currents associated with this magnetic field are non-zero only in 
$r \in [R-\delta, R]$ and can be expressed as
\begin{eqnarray}
\label{corrientes}
j_{\rm r} &=& 0, \nonumber \\
j_\theta &=& - \frac{B_{\rm b} \sin\gamma'_{\rm b}-B_{\rm t}}{\mu \delta} \\
j_{y'} &=& \frac{2 \, B_{\rm b}}{\mu \delta} \, \left[2 -\frac{R-\delta}{r} \right] \, 
\cos \gamma'_{\rm b} \cos\theta, \nonumber
\end{eqnarray}
$\mu$ being the magnetic permeability in vacuum. The current in the
direction of the tube axis, $j_{y'}$, is proportional to the jump of
the vertical component of the magnetic field vector. It reaches a
maximum at $z' = 0$ and vanishes at the top and bottom of the tube
(where $\theta = 90^\circ$ and $\theta = 270^\circ$).

\end{appendix}


\begin{thebibliography}{}

\bibitem[]{} Beck, C.\ 2008, \aap, 480, 825

\bibitem[]{}Bellot Rubio, L.R.\ 2004, Reviews in Modern Astronomy, 17, 21

\bibitem[{{Bellot Rubio}(2007)}]{Bellot:2007} Bellot Rubio, L.R.\
2007, in Highlights of Spanish Astrophysics IV, ed.\ F.\ Figueras et
al.\ (Dordretch: Springer), 271


\bibitem[Bellot Rubio et al.(2004)]{2004A&A...427..319B} Bellot Rubio, 
L.R., Balthasar, H., \& Collados, M.\ 2004, \aap, 427, 319 

\bibitem[]{}Bellot Rubio, L.R., Langhans, K., Schlichenmaier, R.\
2005, \aap, 443, L7

\bibitem[Bellot Rubio et al.(2006)]{2006A&A...453.1117B} Bellot Rubio, L.~R., 
Schlichenmaier, R., \& Tritschler, A.\ 2006, \aap, 453, 1117 

\bibitem[Bellot Rubio et al.(2007)]{2007ApJ...668L..91B} Bellot Rubio, 
L.~R., et al.\ 2007, \apjl, 668, L91 

\bibitem[]{}Borrero, J.M. 2007, \aap, 471, 967

\bibitem[Borrero et al.(2007)]{2007ApJ...666L.133B} Borrero, J.~M., Bellot 
Rubio, L.~R., M\"uller, D.~A.~N.\ 2007, \apjl, 666, L133

\bibitem[Borrero et al.(2005)]{2005A&A...436..333B} Borrero, J.~M., 
Lagg, A., Solanki, S.~K., \& Collados, M.\ 2005, \aap, 436, 333

\bibitem[]{}Borrero, J.M., Rempel, M., \& Solanki, S.K.\ 2006, in  
Solar Polarization IV, ed.\ R.\ Casini, \& B.W.\ Lites, ASP Conf.\ 
Ser., 358, 19

\bibitem[]{}Borrero, J.M., Solanki, S.K., Bellot Rubio, L.R., Lagg, A.,
Mathew, S.K.\ 2004, \aap, 422, 1093

\bibitem[Collados et al.(1994)]{1994A&A...291..622C} Collados, M., 
Mart\'{\i}nez Pillet, V., Ruiz Cobo, B., del Toro Iniesta, J.C., 
\& V\'azquez, M.\ 1994, \aap, 291, 622

\bibitem[Cox \& Giuli(1968)]{1968QB801.C65......} Cox, J.P., \& Giuli, 
R.T.\ 1968, Principles of Stellar Structure, (New York: Gordon and Breach)  

%\bibitem[]{} Evershed, J. 1909, \mnras, 69, 454

\bibitem[Heinemann et al.(2007)]{2007ApJ...669.1390H} Heinemann, T., 
Nordlund, {\AA}., Scharmer, G.~B., \& Spruit, H.~C.\ 2007, \apj, 669, 1390 

\bibitem[]{} Ichimoto, K., et al.\ 2007, PASJ, 59, S593 

\bibitem[]{} Jur\v{c}ak, J., et al.\ 2007, PASJ, 59, S601

\bibitem[Langhans et al.(2007)]{2007A&A...464..763L} Langhans, K., 
Scharmer, G.~B., Kiselman, D., L\"ofdahl, M.~G.\ 2007, \aap, 464, 763 

\bibitem[]{}Levermore, C.D., \& Pomraning, G.C.\  1981, \apj, 248, 321

\bibitem[]{}Kopeck\'y, M. \& Kuklin, G.V.\ 1969, \solphys, 6, 241

\bibitem[]{}Mihalas, D.\ 1978, Stellar Atmospheres, (San Francisco: Freeman and Co.)

\bibitem[]{}Moreno-Insertis, F., Sch\"ussler. M., \& Glampedakis, K.\
2002, \aap, 388, 1022

\bibitem[M{\"u}ller et al.(2002)]{2002A&A...393..305M} M{\"u}ller, 
D.~A.~N., Schlichenmaier, R., Steiner, O., \& Stix, M.\ 2002, \aap, 393, 305

\bibitem[Rimmele(2008)]{2008ApJ...672..684R} Rimmele, T.\ 2008, \apj, 672, 
684 

\bibitem[Rimmele \& Marino(2006)]{2006ApJ...646..593R} 
Rimmele, T., \& Marino, J.\ 2006, \apj, 646, 593 

\bibitem[]{}Rouppe van der Voort, L.H.M., L\"ofdahl, M.G., Kiselman, D.,
Scharmer, G.B. \ 2004, \aap, 414, 717

\bibitem[]{} Sainz Dalda, A., \& Bellot Rubio, L.R. 2008, \aap, 481, L21

\bibitem[Scharmer et al.(2002)]{2002Natur.420..151S} Scharmer, G.B.,
Gudiksen, B.V., Kiselman, D., L\"ofdahl, M.G., \& Rouppe van der
Voort, L.H.M.\ 2002, \nat, 420, 151

\bibitem[]{}Scharmer, G.B., \& Spruit, H.C.\ 2006, \aap, 460, 605

\bibitem[Schlichenmaier(2002)]{2002AN....323..303S} Schlichenmaier, R.\ 
2002, AN, 323, 303 

\bibitem[Schlichenmaier(2003)]{2003ASPC..286..211S} Schlichenmaier, R.\ 
2003, in Current Theoretical Models and Future High Resolution Solar 
Observations: Preparing for ATST, ASP Conf. Ser., 286, 211 

\bibitem[Schlichenmaier \& Schmidt(2000)]{2000A&A...358.1122S} 
Schlichenmaier, R., \& Schmidt, W.\ 2000, \aap, 358, 1122 

\bibitem[]{} Schlichenmaier, R., \& Solanki, S.K.\ 2003, \aap, 411, 257

\bibitem[]{} Schlichenmaier, R., Jahn, K., \& Schmidt, H.U.\ 1998, \aap, 337, 897

\bibitem[]{} Schlichenmaier, R., Bruls, J.H.M.J., \& Sch\"ussler, M.\
1999, \aap, 349, 961

%\bibitem[]{} Schlichenmaier, R., M\"uller, D.A.N., Steiner, O., \& Stix, M.\ 2002, 
%\aap, 381, L77

\bibitem[]{} Sch\"ussler, M., \& V\"ogler, A.\ 2006, \apj, 641, L73

\bibitem[]{} Solanki, S.K.\ 2003, \aapr, 11, 153

\bibitem[]{} Solanki, S.K., \& Montavon C.A.P., \ 1993, \aap, 275, 283

\bibitem[]{} Spruit, H.C.\ 1977, \solphys, 55, 3

\bibitem[]{} Spruit, H.C., \& Scharmer, G.B.\ 2006, \aap, 447, 343 

\bibitem[Stanchfield et al.(1997)]{1997ApJ...477..485S} Stanchfield, 
D.C.H., Thomas, J.H., \& Lites, B.W.\ 1997, \apj, 477, 485 

\bibitem[]{}S\"utterlin, P., Bellot Rubio, L.R., \& Schlichenmaier,
R.\ 2004, \aap, 424, 1049

\bibitem[Title et al.(1993)]{1993ApJ...403..780T} Title, A.M., Frank, 
Z.A., Shine, R.A., Tarbell, T.D., Topka, K.P., Scharmer, G., \& 
Schmidt, W.\ 1993, \apj, 403, 780 

\bibitem[Tritschler et al.(2007)]{2007ApJ...671L..85T} Tritschler, A., 
M{\"u}ller, D.~A.~N., Schlichenmaier, R., \& Hagenaar, H.~J.\ 2007, 
\apjl, 671, L85 

\bibitem[van Noort 
\& Rouppe van der Voort(2008)]{2008arXiv0805.4296V} van Noort, M.J., 
\& Rouppe van der Voort, L.H.M.\ 2008, A\&A, in press, [arXiv:0805.4296] 


\bibitem[Westendorp Plaza et al.(2001)]{2001ApJ...547.1148W}
Westendorp Plaza, C., del Toro Iniesta, J.C., Ruiz Cobo, B., \&
Mart\'{\i}nez Pillet, V.\ 2001, \apj, 547, 1148
\end{thebibliography}
\end{document}